\documentclass[aps,prb,english,twocolumn,showpacs,preprintnumbers,amsmath,amssymb,floatfix]{revtex4}
\usepackage[T1]{fontenc}
\usepackage[latin9]{inputenc}
\usepackage{graphicx}
\usepackage{amssymb}
\usepackage{babel}

\makeatother

\makeatother

\begin{document}

\title{Casimir interaction with an ${1/r}$-dependence}

\author{Bo E. Sernelius}

\affiliation{Division of Theory and Modeling, Department of Physics, Chemistry
and Biology, Link\"{o}ping University, SE-581 83 Link\"{o}ping, Sweden}

\email{bos@ifm.liu.se}

\begin{abstract}
We show that in theory it is possible to obtain a Casimir interaction potential that 
varies with distance as ${1/r}$. We achieve this by invoking hypothetical 
particles having a harmonic oscillator interaction potential. The derivation 
parallels the derivation of the Casimir-Polder interaction between atoms 
in electromagnetism.
\end{abstract}

\maketitle

\section{Introduction}

Johan Diderik van der Waals found by studying the equation of state for
real gases that there is an attractive force between the gas atoms.  The
force is still there even if the atoms have closed electron shells.  He was
awarded the Nobel Price in 1910 for this and related work.  The van der
Waals (vdW) force was found on empirical grounds and it was not until 1930 that
London \cite{Lon,Lon2} gave a realistic explanation for this force.  This
force is of electromagnetic origin and is caused by fluctuations in the
electron density within the atoms.  It is called a dispersion force.  This
force is long range and has the distance dependence, $r^{ - 7}$; the
interaction potential varies as $r^{ - 6}$.  Later Casimir and Polder
\cite{CasPol} realized that if the distance is large enough the finite speed
of light makes the force drop off faster, as $r^{ - 8}$; the interaction
potential varies as $r^{ - 7}$. Both in the van der Waals range and in
the Casimir limit the force is due to fluctuating dipoles.  For smaller
separations also higher order multipole contributions appear and the
distance dependence becomes more complex. The interaction potential can be 
obtained as the sum of the zero-point energy of all the electromagnetic 
normal modes \cite{Ser} of the system. One can divide the modes into two 
categories: one that originates in the atoms; one that originates in the 
surrounding vacuum. The first (second) type dominates in the vdW (Casimir) 
region. The Casimir-Polder potential is obtained \cite{Ser} as
\begin{equation}
\begin{array}{c}
 V_{CP} (r) = - \frac{\hbar }{{\pi r^6 }}\int\limits_0^\infty {d\omega
 \alpha _1 \left( {i\omega } \right)} \alpha _2 \left( {i\omega }
 \right)e^{{{ - 2\omega r} \mathord{\left/ {\vphantom {{ - 2\omega r} c}}
 \right.  \kern-\nulldelimiterspace} c}} \\
 \,\,\,\,\,\,\,\,\,\,\,\,\,\,\,\,\,\,\,\,\,\,\,\left[ {3 + 6\left(
 {{{\omega r} \mathord{\left/ {\vphantom {{\omega r} c}} \right. 
 \kern-\nulldelimiterspace} c}} \right) + 5\left( {{{\omega r}
 \mathord{\left/ {\vphantom {{\omega r} c}} \right. 
 \kern-\nulldelimiterspace} c}} \right)^2 } \right.  \\
 \,\,\,\,\,\,\,\,\,\,\,\,\,\,\,\,\,\,\,\,\,\,\,\,\,\,\,\,\,\,\,\,\,\,\,\,\left.
  { + 2\left( {{{\omega r} \mathord{\left/ {\vphantom {{\omega r} c}}
 \right.  \kern-\nulldelimiterspace} c}} \right)^3 + \left( {{{\omega r}
 \mathord{\left/ {\vphantom {{\omega r} c}} \right. 
 \kern-\nulldelimiterspace} c}} \right)^4 } \right], \\ \end{array}
\end{equation}
where $\alpha _j \left( \omega  \right),\,\,j = 1,2$ is the polarizability 
of atom $j$. This expression is valid in both the vdW and Casimir ranges. 
In the vdW region we have
\begin{equation}
\begin{array}{l}
 V_{vdW} (r)\mathop \approx \limits_{{{\omega r} \mathord{\left/ {\vphantom
 {{\omega r} c}} \right.  \kern-\nulldelimiterspace} c} \to 0} -
 \frac{{3\hbar }}{\pi }\frac{1}{{r^6 }}\int\limits_0^\infty {d\omega \alpha
 _1 \left( {i\omega } \right)} \alpha _2 \left( {i\omega } \right) \\
 \,\,\,\,\,\,\,\,\,\,\,\,\,\,\,\,\,\,\,\,\,\, = - {\textstyle{{3\hbar }
 \over 2}}{\textstyle{{\alpha _1 \left( 0 \right)\alpha _2 \left( 0
 \right)} \over {r^6 }}}\frac{{\omega _1 \omega _2 }}{{\omega _1 + \omega
 _2 }}, \\ \end{array}
\end{equation}
where we in the last step have used the so-called London approximation 
\cite{Lon,Lon2} for the 
 atomic polarizabilities, $\alpha _j \left( {i\omega } \right) \approx
 {{\alpha _j \left( 0 \right)} \mathord{\left/ {\vphantom {{\alpha _j
 \left( 0 \right)} {\left[ {1 + \left( {{\omega \mathord{\left/ {\vphantom
 {\omega {\omega _ij}}} \right.  \kern-\nulldelimiterspace} {\omega _j }}}
 \right)^2 } \right]}}} \right.  \kern-\nulldelimiterspace} {\left[ {1 +
 \left( {{\omega \mathord{\left/ {\vphantom {\omega {\omega _j }}} \right. 
 \kern-\nulldelimiterspace} {\omega _j}}} \right)^2 } \right]}}$. In the 
 Casimir range we have
\begin{equation}
V_C (r)\mathop \approx \limits_{r \to \infty } - \frac{{23\hbar c}}{{4\pi
}}\left[ {\alpha _1 \left( 0 \right)\alpha _2 \left( 0 \right)}
\right]\frac{1}{{r^7 }}.
\end{equation}

Similar forces act between macroscopic objects. Also here they are called 
vdW and Casimir forces. The forces between spherical objects behave just as 
between two atoms and the polarizabilities are now the ones for spheres. 
Finite objects, whatever the shape behave as spheres if the separation is 
large enough. If the separation is not large enough the distance 
dependence of the force depends on the geometrical shapes of the objects 
and the material the objects are made of. It depends on the dispersion of 
the modes in the objects and in the medium surrounding the objects.

In Casimir's famous work \cite{Cas} from 1948 he studied an idealized 
system of two parallel perfectly reflecting metal plates. In this 
situation there are no metal bulk modes or metal surface modes involved at 
all; only vacuum modes contribute to the interaction and force. This means 
that there are no vdW range. The Casimir result is valid at all 
separations, $d$, between the plates. He found:
\begin{equation}
V_C (r) =  - \frac{{\hbar c\pi ^2 }}{{720}}\frac{1}{{d^3 }},
\end{equation}
for the interaction potential and
\begin{equation}
F_C (r) =  - \frac{{\hbar c\pi ^2 }}{{240}}\frac{1}{{d^4 }}
\end{equation}
for the force.  In vacuum there are only transverse modes with dispersion
$\omega = cq$.  If the vacuum is replaced by a purely dielectric medium
there are still only transverse modes but their dispersion is different,
$\omega = \tilde cq$, where ${\tilde c}$ is the speed of light in the
medium.  The only effect on the interaction potential and force is that the
speed of light in vacuum is replaced by the speed of light in the medium. 
However, the purely dielectric medium is an idealization.  In any real
medium there are both transverse and longitudinal modes and the force
becomes more complex.  If the idealized perfectly reflecting metal plates
in Casimir's thought experiment is replaced by plates of real metals there
is a vdW range for smaller distances, smaller than approximately ${c
\mathord{\left/ {\vphantom {c {\omega _{pl} }}} \right. 
\kern-\nulldelimiterspace} {\omega _{pl} }}$, where ${\omega _{pl} }$ is
the plasma frequency of the metals.  In this range the interaction
potential \cite{BosSer} (force) varies as $d^{ - 2}$ ($d^{ - 3}$ ).

A related problem where we clearly see the effect of the dispersion of the 
normal modes of the objects is two parallel 2D (two-dimensional) metal sheets. 
For a 2D metal sheet there are no bulk modes or surface modes. 
The modes are 2D plasmons. These have a dispersion $\omega  \sim \sqrt 
q $. This leads to a fractional separation dependence in the vdW range 
\cite{SerBjo,BosSer,Ser}; the interaction potential (force) varies as $d^{ - {5
\mathord{\left/
 {\vphantom {5 2}} \right. \kern-\nulldelimiterspace} 2}} $($d^{ - {7
 \mathord{\left/ {\vphantom {7 2}} \right.  \kern-\nulldelimiterspace} 2}}
 $).  In the Casimir limit the interaction is the same as for two perfectly
 reflecting plates.  If the dispersion of the normal modes of the sheets were
 different the forces would have a different separation dependence in the 
 vdW range.

To summarize, there are different ways to influence the separation 
dependence of the dispersion forces between objects. It depends on the 
shape of the objects, the material the objects are made from and the 
properties of the surrounding medium. In this work we are more drastic and 
study a system where the fundamental interaction is not electromagnetic. We 
want to see if it is possible to obtain an interaction potential that 
varies with distance as ${1/r}$, like the fundamental interaction 
potentials in electromagnetism and gravity. In Sec. \ref {Basis} we 
introduce the basic assumptions, the formalism and notation. Sec. 
\ref{Dipole} is devoted to the derivation of the fields from a dipole. 
These results are used as the starting point for the derivation of the 
dispersion potential and force in Sec. \ref{DisPot}. Finally, in Sec. 
\ref{Sum} follows summary and discussions.

\section{Theoretical Basis}\label{Basis}

The dispersion forces between atoms have an electromagnetic origin where
the fundamental interaction is the Coulomb interaction between charged
particles, electrons and protons.  We will in this work introduce
hypothetical particles with a different fundamental interaction potential. 
We make the basic assumptions that this interaction travels in vacuum with
the speed of light and that Einstein's two postulates in special relativity
holds also for this interaction.  We will then parallel the derivation of
the dispersion forces in electromagnetism.

We have reasons to believe that any fundamental interaction
potential should form closed classical particle orbits.  There are only two
central force fields where all bound orbits are closed: one has the
interaction potential $V\sim- r^{-1} $ and the other has $V\sim r^2$.  The
second type is known as a harmonic oscillator potential and is our obvious
choice of candidate.

Our notation throughout this work is in analogy with the electromagnetic
case.  We put a tilde above the quantities to distinguish them from the
electromagnetic counterparts.  We assume that the particles have charge
$\pm \tilde q$ and postulate that the electric field from a charge $\tilde
q$ at the origin is ${\bf{\tilde E}} = \tilde q{\bf{r}}$.  This gives the scalar
potential the form $\tilde
\varphi \left( r \right) = - \tilde q{{r^2 } \mathord{\left/ {\vphantom
{{r^2 } 2}} \right.  \kern-\nulldelimiterspace} 2}$. This potential is 
without bounds. This means that the particles can
never be found individually; they are always in pairs.  It takes an
infinite energy to separate them completely.  If we assume that they are
each other's anti-particles the pair will turn into two pairs when one
tries to separate them past a critical distance. This type of
interaction leads to composite particles that are neutral and have no
direct interaction; dispersion interactions due to fluctuations are
possible.   

To find the effect from fluctuating electric dipoles we have to start with
the field from a dipole with dipole moment ${\bf{\tilde p}} = \tilde q{\bf{d}}$.  It
is found to be ${\bf{\tilde E}}\left( {\bf{r}} \right) = - {\bf{\tilde
p}}$.  Thus the field from a dipole is just minus the dipole moment.  There
are no quadrupole or higher order multi-pole contributions as opposed to in
the ordinary electromagnetic theory.  Furthermore the field has no spatial
dependence.  This holds for any static distribution of charges within the
composite particle -- the field lacks spatial dependence.  This implicates
that the only static field is a dipole field and this dipole field is
constant throughout all space, independent of the position of the
particles.  There are no other multi-pole fields.  This is very
encouraging.

From Einstein's two postulates in special relativity follows that there has
to be a magnetic field companion to the electric field.  Furthermore these
two fields obey the two homogeneous Maxwell's equations, $\nabla \times
{\bf{\tilde E}} + \left( {{1 \mathord{\left/ {\vphantom {1 c}} \right. 
\kern-\nulldelimiterspace} c}} \right){{\partial {\bf{\tilde B}}}
\mathord{\left/ {\vphantom {{\partial {\bf{\tilde B}}} {\partial t}}}
\right.  \kern-\nulldelimiterspace} {\partial t}} = 0$ and $\nabla \cdot
{\bf{\tilde B}} = 0$.  This means that we may introduce scalar and vector
potentials, $\tilde \varphi $ and ${\bf{\tilde A}}$, respectively, where
${\bf{\tilde B}} = \nabla \times {\bf{\tilde A}}$ and ${\bf{\tilde E}} = -
\nabla \tilde \varphi - {{\left( {{1 \mathord{\left/ {\vphantom {1 c}}
\right.  \kern-\nulldelimiterspace} c}} \right)\partial {\bf{\tilde A}}}
\mathord{\left/ {\vphantom {{\left( {{1 \mathord{\left/ {\vphantom {1 c}}
\right.  \kern-\nulldelimiterspace} c}} \right)\partial {\bf{\tilde A}}}
{\partial t}}} \right.  \kern-\nulldelimiterspace} {\partial t}}$, all in
complete analogy with in electromagnetism. 

Now we have all we need for our derivations. We start by 
determining, in the next section, the fields from a time-dependent electric 
dipole.

\section{Fields from a Time-Dependent Dipole}\label{Dipole}

In this section we derive the fields from a time-dependent dipole.  We
perform the derivation along the lines used by Heald and Marion
\cite{Heal} on a Hertzian dipole at the origin.  We start with the retarded
potentials,
\begin{eqnarray}
\begin{array}{l}
 \tilde \varphi \left( {{\bf{r}},t} \right) = - {\textstyle{1 \over 2}}\int
 {d^3 r'\tilde \rho \left( {{\bf{r}}',t - {R \mathord{\left/ {\vphantom {R
 c}} \right.  \kern-\nulldelimiterspace} c}} \right)R^2 } ; \\ {\bf{\tilde
 A}}\left( {{\bf{r}},t} \right) = {\textstyle{1 \over {2c}}}\int {d^3
 r'{\bf{\tilde J}}\left( {{\bf{r}}',t - {R \mathord{\left/ {\vphantom {R
 c}} \right.  \kern-\nulldelimiterspace} c}} \right)R^2 } ;\quad {\bf{R}} =
 {\bf{r}} - {\bf{r}}', \\ 
\end{array}
\end{eqnarray}
where $\tilde \rho$ and ${\bf{\tilde J}}$ are the charge and current
densities, respectively. From the relations
\begin{equation}
\begin{array}{l}
 {\bf{{\tilde B}}} = \nabla \times {\bf{{\tilde A}}} \\ {\bf{{\tilde E}}} = - 
 \nabla {\tilde \varphi }-
 {{{\textstyle{1 \over c}}\partial {\bf{{\tilde A}}}} \mathord{\left/ {\vphantom
 {{{\textstyle{1 \over c}}\partial {\bf{{\tilde A}}}} {\partial t}}} \right. 
 \kern-\nulldelimiterspace} {\partial t}} \\ \end{array}
\end{equation}
follow
\begin{equation}
\begin{array}{l}
 {\bf{{\tilde E}}}\left( {{\bf{r}},t} \right) = \int\limits_V {\left( {\left[ 
 {\tilde \rho}
 \right]R\hat R - {\textstyle{1 \over {2c}}}\left[ {\dot {\tilde \rho} } \right]R^2
 \hat R - \left[ {{\bf{{  \dot {\tilde J}}}}} \right]{\textstyle{{R^2 } \over {2c^2 }}}}
 \right)dv'} \\ \,\,\,\,\,\,\,\,\,\,\,\,\,\,\,\,\, = {\bf{{\tilde E}}}_1 \left(
 {{\bf{r}},t} \right) + {\bf{{\tilde E}}}_2 \left( {{\bf{r}},t} \right) + {\bf{{\tilde E}}}_3
 \left( {{\bf{r}},t} \right); \\ {\bf{{\tilde B}}}\left( {{\bf{r}},t} \right) =
 \int\limits_V {\left( { - \left[ {\bf{{\tilde J}}} \right] \times \hat
 R{\textstyle{R \over c}} + \left[ {{\bf{ \dot{\tilde J}}}} \right] \times \hat
 R{\textstyle{{R^2 } \over {2c^2 }}}} \right)dv'} \\
 \,\,\,\,\,\,\,\,\,\,\,\,\,\,\,\,\, = {\bf{{\tilde B}}}_1 \left( {{\bf{r}},t}
 \right) + {\bf{{\tilde B}}}_2 \left( {{\bf{r}},t} \right), \\ \end{array}
\end{equation}
where a dot means the time derivative and square brackets that the function
within the brackets is determined at retarded times, $t - {r
\mathord{\left/ {\vphantom {R c}} \right.  \kern-\nulldelimiterspace} c}$. 
We have chosen a spherical coordinate system with its $z$-axis along {\bf
p}.  We have used the relations
\begin{equation}
\begin{array}{l}
 \nabla \left( {\left[{\tilde \rho} \right]R^2 } \right) = R^2 \nabla \left[ {\tilde \rho}
 \right] + \left[{\tilde \rho} \right]\nabla \left( {R^2 } \right); \\ \nabla
 \left[ {\tilde \rho} \right] = \left[ {\dot {\tilde \rho} } \right]\nabla \left( {t -
 {{\left| {{\bf{r}} - {\bf{r}}'} \right|} \mathord{\left/ {\vphantom
 {{\left| {{\bf{r}} - {\bf{r}}'} \right|} c}} \right. 
 \kern-\nulldelimiterspace} c}} \right) = - {\textstyle{1 \over c}}\left[
 {\dot {\tilde \rho} } \right]\hat R; \\ \nabla \times \left( {\left[ 
 {\bf{{\tilde J}}}
 \right]R^2 } \right) = R^2 \nabla \times \left[ {\bf{{\tilde J}}} \right] - \left[
 {\bf{{\tilde J}}} \right] \times \nabla \left( {R^2 } \right); \\ \nabla \times
 \left[ {\bf{{\tilde J}}} \right] = - \left[ {{\bf{\dot {\tilde J}}}} \right] \times \nabla
 \left( {t - {{\left| {{\bf{r}} - {\bf{r}}'} \right|} \mathord{\left/
 {\vphantom {{\left| {{\bf{r}} - {\bf{r}}'} \right|} c}} \right. 
 \kern-\nulldelimiterspace} c}} \right) = {\textstyle{1 \over c}}\left[
 {{\bf{\dot {\tilde J}}}} \right] \times \hat R. \\ \end{array}
\end{equation}
Now, we apply these relations to a Hertzian dipole at the origin pointing 
in the ${\hat z}$-direction. We have

\begin{equation}
{\bf{{\tilde J}}}d{v}' \to {\tilde I}dt\hat z \to {\textstyle{{\dot {\tilde p}} \over d}}{\bf{d}}
= {\bf{\dot {\tilde p}}}\;{\rm{and}}\;R \to r.
\end{equation}
This gives 
\begin{equation}
\begin{array}{l}
 {\bf{\tilde E}}_3 = - {\textstyle{1 \over 2}}\left( {{\textstyle{r \over
 c}}} \right)^2 \left[ {{\bf{\ddot{ \tilde p}}}} \right]; \\ {\bf{\tilde
 B}}_1 = - \left( {{\textstyle{r \over c}}} \right)\left[ {{\bf{\dot {\tilde
 p}}}} \right] \times \hat r; \\ {\bf{\tilde B}}_2 = {\textstyle{1 \over
 2}}\left( {{\textstyle{r \over c}}} \right)^2 \left[ {{\bf{\ddot{ \tilde
 p}}}} \right] \times \hat r. \\ \end{array}
\end{equation} 
 We study the two remaining fields on the axis of the dipole and in the
 equatorial plane.  From these we get the fields in a general point.  We
 start with the axis and let $\Delta t = {d \mathord{\left/ {\vphantom {d
 {2c}}} \right.  \kern-\nulldelimiterspace} {2c}}$,

\begin{equation}
\begin{array}{l}
 \left( {{\bf{\tilde E}}_1 } \right)_{ax} = \left[ {\tilde q\left( {t' +
 \Delta t} \right)\left( {z - {d \mathord{\left/ {\vphantom {d 2}} \right. 
 \kern-\nulldelimiterspace} 2}} \right)} \right.  \\ \left. 
 {\,\,\,\,\,\,\,\,\,\,\,\,\,\,\,\,\,\,\,\,\,\,\,\,\,\,\,\,\,\, - \tilde
 q\left( {t' - \Delta t} \right)\left( {z + {d \mathord{\left/ {\vphantom
 {d 2}} \right.  \kern-\nulldelimiterspace} 2}} \right)} \right]\hat z \\
 \quad \quad \;\; = \left[ {\left( {\left[ {\tilde q} \right] + {{\left[
 {\dot {\tilde q}} \right]d} \mathord{\left/ {\vphantom {{\left[ {\dot
 {\tilde q}} \right]d} {2c}}} \right.  \kern-\nulldelimiterspace} {2c}}}
 \right)\left( {z - {d \mathord{\left/ {\vphantom {d 2}} \right. 
 \kern-\nulldelimiterspace} 2}} \right)} \right.  \\
 \,\,\,\,\,\,\,\,\,\,\,\,\,\,\,\,\,\,\,\,\,\,\left.  { - \left( {\left[
 {\tilde q} \right] - {{\left[ {\dot {\tilde q}} \right]d} \mathord{\left/
 {\vphantom {{\left[ {\dot {\tilde q}} \right]d} {2c}}} \right. 
 \kern-\nulldelimiterspace} {2c}}} \right)\left( {z + {d \mathord{\left/
 {\vphantom {d 2}} \right.  \kern-\nulldelimiterspace} 2}} \right)}
 \right]\hat z \\ \quad \quad \;\; = \left[ {\left( { - d\left[ {\tilde q}
 \right] + {{\left[ {\dot {\tilde q}} \right]zd} \mathord{\left/ {\vphantom
 {{\left[ {\dot {\tilde q}} \right]zd} c}} \right. 
 \kern-\nulldelimiterspace} c}} \right)} \right]\hat z \\ \quad \quad \;\;
 = - \left[ {{\bf{\tilde p}}} \right] + \left[ {{\bf{\dot {\tilde p}}}}
 \right]{\textstyle{z \over c}}, \\ \end{array}
\end{equation}
and
\begin{equation}
\begin{array}{l}
 \left( {{\bf{\tilde E}}_2 } \right)_{ax} = - {\textstyle{1 \over
 {2c}}}\left[ {\dot {\tilde q}\left( {t' + \Delta t} \right)\left( {z - {d
 \mathord{\left/ {\vphantom {d 2}} \right.  \kern-\nulldelimiterspace} 2}}
 \right)^2 } \right.  \\ \quad \quad \quad \quad \left.  { - \dot {\tilde
 q}\left( {t' - \Delta t} \right)\left( {z + {d \mathord{\left/ {\vphantom
 {d 2}} \right.  \kern-\nulldelimiterspace} 2}} \right)^2 } \right]\hat z
 \\ \quad \quad \quad = - {\textstyle{1 \over {2c}}}\left[ {\left( {\left[
 {\dot {\tilde q}} \right] + \left[ {\ddot {\tilde q}} \right]{d
 \mathord{\left/ {\vphantom {d {2c}}} \right.  \kern-\nulldelimiterspace}
 {2c}}} \right)\left( {z^2 - dz} \right)} \right.  \\ \quad \quad \quad
 \quad \left.  { - \left( {\left[ {\dot {\tilde q}} \right] - \left[ {\ddot
{\tilde q}} \right]{d \mathord{\left/ {\vphantom {d {2c}}} \right. 
 \kern-\nulldelimiterspace} {2c}}} \right)\left( {z^2 + dz} \right)}
 \right]\hat z \\ \quad \quad \quad = - {\textstyle{1 \over {2c}}}\left[
 {\left[ {\dot {\tilde q}} \right]\left( { - 2dz} \right) + \left[ {\ddot
 {\tilde q}} \right]{{dz^2 } \mathord{\left/ {\vphantom {{dz^2 } c}} \right. 
 \kern-\nulldelimiterspace} c}} \right]\hat z \\ \quad \quad \quad = \left[
 {{\bf{\dot {\tilde q}}}} \right]{\textstyle{z \over c}} - {\textstyle{1
 \over 2}}\left[ {{\bf{\ddot {\tilde p}}}} \right]\left( {{\textstyle{z \over
 c}}} \right)^2 .\\ \end{array}
\end{equation}
The fields in the equatorial plane are
\begin{equation}
\begin{array}{l}
 \left( {{\bf{\tilde E}}_1 } \right)_{eq} = \left[ {\left[ {\tilde q}
 \right]\sqrt {r^2 + \left( {{d \mathord{\left/ {\vphantom {d 2}} \right. 
 \kern-\nulldelimiterspace} 2}} \right)^2 } 2{{\left( {{d \mathord{\left/
 {\vphantom {d 2}} \right.  \kern-\nulldelimiterspace} 2}} \right)}
 \mathord{\left/ {\vphantom {{\left( {{d \mathord{\left/ {\vphantom {d 2}}
 \right.  \kern-\nulldelimiterspace} 2}} \right)} {\sqrt {r^2 + \left( {{d
 \mathord{\left/ {\vphantom {d 2}} \right.  \kern-\nulldelimiterspace} 2}}
 \right)^2 } }}} \right.  \kern-\nulldelimiterspace} {\sqrt {r^2 + \left(
 {{d \mathord{\left/ {\vphantom {d 2}} \right.  \kern-\nulldelimiterspace}
 2}} \right)^2 } }}} \right]\hat \theta \\ \quad \quad \quad = - \left[
 {{\bf{\tilde p}}} \right], \\ \end{array}
\end{equation}
and
\begin{equation}
\begin{array}{l}
 \left( {{\bf{\tilde E}}_2 } \right)_{eq} = - {\textstyle{1 \over
 {2c}}}\left[ {\left[ {\dot {\tilde q}} \right]\left[ {r^2 + \left( {{d
 \mathord{\left/ {\vphantom {d 2}} \right.  \kern-\nulldelimiterspace} 2}}
 \right)^2 } \right]} \right.  \\ \quad \quad \quad \quad \left. 
 {2{{\left( {{d \mathord{\left/ {\vphantom {d 2}} \right. 
 \kern-\nulldelimiterspace} 2}} \right)} \mathord{\left/ {\vphantom
 {{\left( {{d \mathord{\left/ {\vphantom {d 2}} \right. 
 \kern-\nulldelimiterspace} 2}} \right)} {\sqrt {r^2 + \left( {{d
 \mathord{\left/ {\vphantom {d 2}} \right.  \kern-\nulldelimiterspace} 2}}
 \right)^2 } }}} \right.  \kern-\nulldelimiterspace} {\sqrt {r^2 + \left(
 {{d \mathord{\left/ {\vphantom {d 2}} \right.  \kern-\nulldelimiterspace}
 2}} \right)^2 } }}} \right]\hat \theta \to {\textstyle{r \over
 {2c}}}\left[ {{\bf{\dot {\tilde p}}}} \right]. \\ \end{array}
\end{equation}
Now we get
\begin{equation}
\begin{array}{l}
 {\bf{\tilde E}}_1 = \left( { - \left[ {\tilde p} \right] + {\textstyle{r
 \over c}}\left[ {\dot {\tilde p}} \right]} \right)\cos \theta \hat r +
 \left[ {\dot {\tilde p}} \right]\sin \theta \hat \theta ; \\ {\bf{\tilde
 E}}_2 = \left( {{\textstyle{r \over c}}\left[ {\dot {\tilde p}} \right] -
 {\textstyle{1 \over 2}}\left( {{\textstyle{r \over c}}} \right)^2 \left[
 {\ddot {\tilde p}} \right]} \right)\cos \theta \hat r - {\textstyle{1
 \over 2}}\left( {{\textstyle{r \over c}}} \right)\left[ {\dot {\tilde p}}
 \right]\sin \theta \hat \theta ; \\ {\bf{\tilde E}}_3 = - {\textstyle{1
 \over 2}}\left( {{\textstyle{r \over c}}} \right)^2 \left[ {\ddot {\tilde
 p}} \right]\cos \theta \hat r + {\textstyle{1 \over 2}}\left(
 {{\textstyle{r \over c}}} \right)^2 \left[ {\ddot {\tilde p}} \right]\sin
 \theta \hat \theta ,\\ \end{array}
\end{equation}
and the total fields are
\begin{eqnarray}
\begin{array}{l}
 {\bf{\tilde E}}\left( {{\bf{r}},t} \right) = \left\{ { - \left[ {\tilde p}
 \right] + 2\left( {{\textstyle{r \over c}}} \right)\left[ {\dot{ \tilde p}}
 \right] - \left( {{\textstyle{r \over c}}} \right)^2 \left[ {\ddot{ \tilde
 p}} \right]} \right\}\cos \theta \hat r \\ \quad \quad \quad \quad +
 \left\{ {\left[ {\tilde p} \right] - {\textstyle{1 \over 2}}\left(
 {{\textstyle{r \over c}}} \right)\left[ {\dot{ \tilde p} }\right] +
 {\textstyle{1 \over 2}}\left( {{\textstyle{r \over c}}} \right)^2 \left[
 {\ddot{ \tilde p} }\right]} \right\}\sin \theta \hat \theta ; \\ {\bf{\tilde
 B}}\left( {{\bf{r}},t} \right) = \left\{ { - \left( {{\textstyle{r \over
 c}}} \right)\left[ {\dot{ \tilde p} }\right] + {\textstyle{1 \over 2}}\left(
 {{\textstyle{r \over c}}} \right)^2 \left[ {\ddot{ \tilde p} }\right]}
 \right\}\sin \theta \hat \varphi .\\ \end{array}
\end{eqnarray}
Now we have all we need to calculate the dispersion forces between the 
composite particles. This we do in next section.

\section{Dispersion Potential and Force}\label{DisPot}

We define the polarizability, $\tilde \alpha $, for a composite particle
through: ${\bf{\tilde p}} = \tilde \alpha {\bf{\tilde E}}$.  The dispersion
interaction between two particles, 1 and 2, is found from realizing that a
polarization of particle 1 gives rise to an electric field at the position
of particle 2.  This field polarizes particle 2, which results in a field
at the position of particle 1.  Closing this loop results in self-sustained
fields -- normal modes.  We find these in analogy with the treatment of the
two-atom system in the book: Surface Modes in Physics \cite{Ser}. 

We may reformulate the electric field from a time dependent dipole as
\begin{equation}
\begin{array}{l}
 {\bf{\tilde E}}\left( {{\bf{r}},t} \right) = \left( {\hat p \cdot \hat r}
 \right)\hat r\left[ { - \tilde p\left( {t - {r \mathord{\left/ {\vphantom
 {r c}} \right.  \kern-\nulldelimiterspace} c}} \right) + 2\left(
 {{\textstyle{r \over c}}} \right)\dot{ \tilde p}\left( {t - {r
 \mathord{\left/ {\vphantom {r c}} \right.  \kern-\nulldelimiterspace} c}}
 \right)} \right.  \\ \quad \quad \quad \left.  {\quad - \left(
 {{\textstyle{r \over c}}} \right)^2 \ddot{ \tilde p}\left( {t - {r
 \mathord{\left/ {\vphantom {r c}} \right.  \kern-\nulldelimiterspace} c}}
 \right)} \right] \\ \quad \quad \quad \quad + \left[ {\hat p - \left(
 {\hat p \cdot \hat r} \right)\hat r} \right]\left[ { - \tilde p\left( {t -
 {r \mathord{\left/ {\vphantom {r c}} \right.  \kern-\nulldelimiterspace}
 c}} \right)} \right.  \\ \left.  {\quad \quad \quad \quad + {\textstyle{1
 \over 2}}\left( {{\textstyle{r \over c}}} \right)\dot{ \tilde p}\left( {t
 - {r \mathord{\left/ {\vphantom {r c}} \right.  \kern-\nulldelimiterspace}
 c}} \right) - {\textstyle{1 \over 2}}\left( {{\textstyle{r \over c}}}
 \right)^2 \ddot{ \tilde p}\left( {t - {r \mathord{\left/ {\vphantom {r c}}
 \right.  \kern-\nulldelimiterspace} c}} \right)} \right], \\ \end{array}
\end{equation}
and its Fourier transform with respect to time becomes
\begin{equation}
\begin{array}{l}
 {\bf{\tilde E}}\left( {{\bf{r}},\omega } \right) = \tilde p\left( \omega
 \right)e^{i{{\omega r} \mathord{\left/ {\vphantom {{\omega r} c}} \right. 
 \kern-\nulldelimiterspace} c}} \left\{ {\left( {\hat p \cdot \hat r}
 \right)\hat r\left[ { - 1 + 2\left( {{{ - i\omega r} \mathord{\left/
 {\vphantom {{ - i\omega r} c}} \right.  \kern-\nulldelimiterspace} c}}
 \right)} \right.} \right.  \\ \quad \quad \quad \quad \left.  {\quad \quad
 + \left( {{{\omega r} \mathord{\left/ {\vphantom {{\omega r} c}} \right. 
 \kern-\nulldelimiterspace} c}} \right)^2 } \right] + \left[ {\hat p -
 \left( {\hat p \cdot \hat r} \right)\hat r} \right]\left[ { - 1} \right. 
 \\ \left.  {\quad \quad \quad \quad \quad \quad \quad \left.  { +
 {\textstyle{1 \over 2}}\left( {{{ - i\omega r} \mathord{\left/ {\vphantom
 {{ - i\omega r} c}} \right.  \kern-\nulldelimiterspace} c}} \right) +
 {\textstyle{1 \over 2}}\left( {{{\omega r} \mathord{\left/ {\vphantom
 {{\omega r} c}} \right.  \kern-\nulldelimiterspace} c}} \right)^2 }
 \right]} \right\}.  \\ \end{array}
\end{equation}
On tensor form we may write
\begin{equation}
\begin{array}{l}
 {\bf{\tilde E}}\left( {{\bf{r}},\omega } \right) = e^{i{{\omega r}
 \mathord{\left/ {\vphantom {{\omega r} c}} \right. 
 \kern-\nulldelimiterspace} c}} \left\{ {\tilde{ \tilde \alpha }\left[ { -
 1 + 2\left( {{{ - i\omega r} \mathord{\left/ {\vphantom {{ - i\omega r}
 c}} \right.  \kern-\nulldelimiterspace} c}} \right) + \left( {{{\omega r}
 \mathord{\left/ {\vphantom {{\omega r} c}} \right. 
 \kern-\nulldelimiterspace} c}} \right)^2 } \right]} \right.  \\ \left. 
 {\quad \quad \quad \quad + \tilde{ \tilde \beta }\left[ { - 1 +
 {\textstyle{1 \over 2}}\left( {{{ - i\omega r} \mathord{\left/ {\vphantom
 {{ - i\omega r} c}} \right.  \kern-\nulldelimiterspace} c}} \right) +
 {\textstyle{1 \over 2}}\left( {{{\omega r} \mathord{\left/ {\vphantom
 {{\omega r} c}} \right.  \kern-\nulldelimiterspace} c}} \right)^2 }
 \right]} \right\}{\bf{\tilde p}}\left( \omega \right), \\ \end{array}
\end{equation}
where the tensors are: $\tilde{ \tilde \alpha} = {{r_\mu r_\nu }
\mathord{\left/ {\vphantom {{r_\mu r_\nu } {r^2 }}} \right. 
\kern-\nulldelimiterspace} {r^2 }}$; $ \tilde{ \tilde \beta} = \delta _{\mu
\nu } - {{r_\mu r_\nu } \mathord{\left/ {\vphantom {{r_\mu r_\nu } {r^2 }}}
\right.  \kern-\nulldelimiterspace} {r^2 }} = \tilde{ \tilde I }- \tilde{
\tilde \alpha } $.  We let all tensors have double tildes to distinguish
them from our fields.  Things become very simple if we choose the third
principle axis to point along $\bf {r}$.  Then we have

$\tilde{ \tilde \alpha } = \left( {\begin{array}{*{20}c}
   0 & 0 & 0  \\
   0 & 0 & 0  \\
   0 & 0 & 1  \\
\end{array}} \right);\quad \tilde{ \tilde \beta } = \left( {\begin{array}{*{20}c}
   1 & 0 & 0  \\
   0 & 1 & 0  \\
   0 & 0 & 0  \\
\end{array}} \right),\\$
and $\tilde{ \tilde \alpha }^2 = \tilde{ \tilde \alpha };\quad \tilde{
\tilde \beta }^2 = \tilde{ \tilde \beta };\quad \tilde{ \tilde \alpha
}\cdot \tilde{ \tilde \beta }= 0$.  If we now let
\begin{equation}
\begin{array}{l}
 \tilde{ \tilde \gamma }= e^{i{{\omega r} \mathord{\left/ {\vphantom
 {{\omega r} c}} \right.  \kern-\nulldelimiterspace} c}} \left\{ {\tilde{
 \tilde \alpha }\left[ { - 1 + 2\left( {{{ - i\omega r} \mathord{\left/
 {\vphantom {{ - i\omega r} c}} \right.  \kern-\nulldelimiterspace} c}}
 \right) + \left( {{{\omega r} \mathord{\left/ {\vphantom {{\omega r} c}}
 \right.  \kern-\nulldelimiterspace} c}} \right)^2 } \right]} \right.  \\
 \quad \quad \quad \quad + \left.  {\tilde{ \tilde \beta }\left[ { - 1 +
 {\textstyle{1 \over 2}}\left( {{{ - i\omega r} \mathord{\left/ {\vphantom
 {{ - i\omega r} c}} \right.  \kern-\nulldelimiterspace} c}} \right) +
 {\textstyle{1 \over 2}}\left( {{{\omega r} \mathord{\left/ {\vphantom
 {{\omega r} c}} \right.  \kern-\nulldelimiterspace} c}} \right)^2 }
 \right]} \right\}, \\ \end{array}
\end{equation}
we may set up the self-sustained fields between two polarizable particles as
\begin{equation}
\begin{array}{l}
 {\bf{\tilde p}}_2 \left( \omega \right) = \tilde \alpha _2 \left( \omega
 \right){\bf{\tilde E}}_1 \left( {{\bf{r}}_2 ,\omega } \right) = \tilde
 \alpha _2 \left( \omega \right)\tilde{ \tilde \gamma }{\bf{\tilde p}}_1
 \left( \omega \right); \\ {\bf{\tilde p}}_1 \left( \omega \right) = \tilde
 \alpha _1 \left( \omega \right){\bf{\tilde E}}_2 \left( {{\bf{r}}_1
 ,\omega } \right) = \tilde \alpha _1 \left( \omega \right)\tilde{ \tilde
 \gamma }{\bf{\tilde p}}_2 \left( \omega \right), \\ \end{array}
\end{equation}
where ${\bf{\tilde E}}_i \left( {{\bf{r}}_j ,\omega } \right)$ is the
electric field at the position of particle $\it{j}$ caused by the polarized
particle $\it{i}$.  Eliminating ${\bf{p}}_2 $ gives
\begin{equation}
{\bf{\tilde p}}_1 \left( \omega \right) = \tilde \alpha _1 \left( \omega
\right)\tilde{ \tilde \gamma }\tilde \alpha _2 \left( \omega \right)\tilde{
\tilde \gamma }{\bf{\tilde p}}_1 \left( \omega \right) = \tilde \alpha _1
\left( \omega \right)\tilde \alpha _2 \left( \omega \right)\tilde{ \tilde
\gamma }^2 {\bf{\tilde p}}_1 \left( \omega \right),
\end{equation}
where in the last step we have assumed that the polarizabilities are
isotropic.  Now, we have
\begin{equation}
\tilde{ \tilde A}{\bf{\tilde p}}_1 \left( \omega \right) = \left( {\tilde{
\tilde I }- \tilde \alpha _1 \left( \omega \right)\tilde \alpha _2 \left(
\omega \right)\tilde{ \tilde \gamma }^2 } \right){\bf{\tilde p}}_1 \left(
\omega \right) = 0.
\end{equation}
This system of equations has non-trivial solutions, the normal modes, if
$\left| {\tilde{ \tilde A}} \right| = 0$.  Now,
\begin{equation}
\begin{array}{l}
 \tilde{ \tilde \gamma }^2 = e^{i{{2\omega r} \mathord{\left/ {\vphantom
 {{2\omega r} c}} \right.  \kern-\nulldelimiterspace} c}} \left\{ {\tilde{
 \tilde \alpha }\left[ { - 1 + 2\left( {{{ - i\omega r} \mathord{\left/
 {\vphantom {{ - i\omega r} c}} \right.  \kern-\nulldelimiterspace} c}}
 \right) + \left( {{{\omega r} \mathord{\left/ {\vphantom {{\omega r} c}}
 \right.  \kern-\nulldelimiterspace} c}} \right)^2 } \right]} \right.^2 \\
 \quad \quad \quad \quad \left.  { + \tilde{ \tilde \beta }\left[ { - 1 +
 {\textstyle{1 \over 2}}\left( {{{ - i\omega r} \mathord{\left/ {\vphantom
 {{ - i\omega r} c}} \right.  \kern-\nulldelimiterspace} c}} \right) +
 {\textstyle{1 \over 2}}\left( {{{\omega r} \mathord{\left/ {\vphantom
 {{\omega r} c}} \right.  \kern-\nulldelimiterspace} c}} \right)^2 }
 \right]^2 } \right\}, \\ \end{array}
\end{equation}
and
\begin{equation}
\begin{array}{l}
 \left| {\tilde{ \tilde A}\left( \omega \right)} \right| = \left\{ {1 -
 \tilde \alpha _1 \left( \omega \right)\tilde \alpha _2 \left( \omega
 \right)e^{i{{2\omega r} \mathord{\left/ {\vphantom {{2\omega r} c}}
 \right.  \kern-\nulldelimiterspace} c}} \left[ { - 1 + 2\left( {{{ -
 i\omega r} \mathord{\left/ {\vphantom {{ - i\omega r} c}} \right. 
 \kern-\nulldelimiterspace} c}} \right)} \right.} \right.  \\ \quad \quad
 \quad \quad + \left.  {\left.  {\left( {{{\omega r} \mathord{\left/
 {\vphantom {{\omega r} c}} \right.  \kern-\nulldelimiterspace} c}}
 \right)^2 } \right]^2 } \right\} \times \left\{ {1 - \tilde \alpha _1
 \left( \omega \right)\tilde \alpha _2 \left( \omega \right)e^{i{{2\omega
 r} \mathord{\left/ {\vphantom {{2\omega r} c}} \right. 
 \kern-\nulldelimiterspace} c}} } \right.  \\ \quad \quad \quad \quad \quad
 \quad \times \left.  {\left[ { - 1 + {\textstyle{1 \over 2}}\left( {{{ -
 i\omega r} \mathord{\left/ {\vphantom {{ - i\omega r} c}} \right. 
 \kern-\nulldelimiterspace} c}} \right) + {\textstyle{1 \over 2}}\left(
 {{{\omega r} \mathord{\left/ {\vphantom {{\omega r} c}} \right. 
 \kern-\nulldelimiterspace} c}} \right)^2 } \right]^2 } \right\}^2 .  \\
 \end{array}
\end{equation}

We will need the results on the imaginary frequency axis, i.e.,
\begin{equation}
\begin{array}{l}
 \left| {\tilde{ \tilde A}\left( {i\omega } \right)} \right| = \left\{ {1 -
 \tilde \alpha _1 \left( {i\omega } \right)\tilde \alpha _2 \left( {i\omega
 } \right)e^{ - {{2\left| \omega \right|r} \mathord{\left/ {\vphantom
 {{2\left| \omega \right|r} c}} \right.  \kern-\nulldelimiterspace} c}} }
 \right.  \\ \quad \quad \quad \quad \quad \quad \times \left[ { - 1 +
 2\left( {{{\left| \omega \right|r} \mathord{\left/ {\vphantom {{\left|
 \omega \right|r} c}} \right.  \kern-\nulldelimiterspace} c}} \right) -
 \left( {{{\omega r} \mathord{\left/ {\vphantom {{\omega r} c}} \right. 
 \kern-\nulldelimiterspace} c}} \right)^2 } \right]^2 \\
 \,\,\,\,\,\,\,\,\,\,\,\,\,\,\,\,\,\,\,\,\,\,\,\,\,\,\, \times \left\{ {1 -
 \tilde \alpha _1 \left( {i\omega } \right)\tilde \alpha _2 \left( {i\omega
 } \right)e^{ - {{2\left| \omega \right|r} \mathord{\left/ {\vphantom
 {{2\left| \omega \right|r} c}} \right.  \kern-\nulldelimiterspace} c}} }
 \right.  \\ \quad \quad \quad \quad \quad \times \left.  {\left[ { - 1 +
 {\textstyle{1 \over 2}}\left( {{{\left| \omega \right|r} \mathord{\left/
 {\vphantom {{\left| \omega \right|r} c}} \right. 
 \kern-\nulldelimiterspace} c}} \right) - {\textstyle{1 \over 2}}\left(
 {{{\omega r} \mathord{\left/ {\vphantom {{\omega r} c}} \right. 
 \kern-\nulldelimiterspace} c}} \right)^2 } \right]^2 } \right\}^2 .  \\
 \end{array}
\end{equation}
Now, the zero-point energy for the system of two interacting particles, $r$
apart, relative the energy when they are at infinite separation is
\begin{equation}
V\left( r \right) = \frac{1}{{2\pi i}}\oint {dz\left( {\frac{{\hbar z}}{2}}
\right)\frac{d}{{dz}}\ln \left| {\tilde{ \tilde A}\left( z \right)}
\right|},
\end{equation}
where the integration is performed in the complex frequency plane around a
contour enclosing the positive part of the real frequency axis.  This
equation is obtained from the generalized argument principle \cite{Ser}. 
With a standard procedure \cite{Ser} one arrives at an integral along the
imaginary frequency axis, $V\left( r \right) = \left( {{\hbar
\mathord{\left/ {\vphantom {\hbar {2\pi }}} \right. 
\kern-\nulldelimiterspace} {2\pi }}} \right)\int_0^\infty {d\omega \ln
\left| {\tilde{ \tilde A}\left( {i\omega } \right)} \right|} $, and the
resulting interaction potential is
\begin{equation}
\begin{array}{l}
 V\left( r \right) = \frac{\hbar }{{2\pi }}\int\limits_0^\infty {d\omega
 \ln \left\{ {\left[ {1 - \tilde \alpha _1 \left( {i\omega } \right)\tilde
 \alpha _2 \left( {i\omega } \right)e^{ - {{2\omega r} \mathord{\left/
 {\vphantom {{2\omega r} c}} \right.  \kern-\nulldelimiterspace} c}} }
 \right.} \right.} \\ \quad \quad \quad \quad \quad \quad \times \left. 
 {\left[ { - 1 + 2\left( {{{\omega r} \mathord{\left/ {\vphantom {{\omega
 r} c}} \right.  \kern-\nulldelimiterspace} c}} \right) - \left( {{{\omega
 r} \mathord{\left/ {\vphantom {{\omega r} c}} \right. 
 \kern-\nulldelimiterspace} c}} \right)^2 } \right]^2 } \right] \\
 \,\,\,\,\,\,\,\,\,\,\,\,\,\,\,\,\,\,\,\,\,\,\,\,\,\,\,\,\,\,\,\,\,\,\,\,\,\,\,\,
 \times \left[ {1 - \tilde \alpha _1 \left( {i\omega } \right)\tilde \alpha
 _2 \left( {i\omega } \right)e^{ - {{2\omega r} \mathord{\left/ {\vphantom
 {{2\omega r} c}} \right.  \kern-\nulldelimiterspace} c}} } \right.  \\
 \quad \quad \quad \quad \quad \times \left.  {\left.  {\left[ { - 1 +
 {\textstyle{1 \over 2}}\left( {{{\omega r} \mathord{\left/ {\vphantom
 {{\omega r} c}} \right.  \kern-\nulldelimiterspace} c}} \right) -
 {\textstyle{1 \over 2}}\left( {{{\omega r} \mathord{\left/ {\vphantom
 {{\omega r} c}} \right.  \kern-\nulldelimiterspace} c}} \right)^2 }
 \right]^2 } \right]^2 } \right\}.  \\ \end{array}
\end{equation}
Expanding the logarithm, assuming that the polarizabilities are very small,
gives 
\begin{equation}
\begin{array}{l}
 V\left( r \right) = - \frac{\hbar }{{4\pi }}\int\limits_0^\infty {d\omega
 \tilde \alpha _1 \left( {i\omega } \right)\tilde \alpha _2 \left( {i\omega
 } \right)e^{ - {{2\omega r} \mathord{\left/ {\vphantom {{2\omega r} c}}
 \right.  \kern-\nulldelimiterspace} c}} } \left[ {6 - 12\left( {{{\omega
 r} \mathord{\left/ {\vphantom {{\omega r} c}} \right. 
 \kern-\nulldelimiterspace} c}} \right)} \right.  \\ \,\left.  {\quad \quad
 \quad \quad \quad \quad + 17\left( {{{\omega r} \mathord{\left/ {\vphantom
 {{\omega r} c}} \right.  \kern-\nulldelimiterspace} c}} \right)^2 -
 9\left( {{{\omega r} \mathord{\left/ {\vphantom {{\omega r} c}} \right. 
 \kern-\nulldelimiterspace} c}} \right)^3 + 3\left( {{{\omega r}
 \mathord{\left/ {\vphantom {{\omega r} c}} \right. 
 \kern-\nulldelimiterspace} c}} \right)^4 } \right].  \\ \end{array}
\end{equation}

Two separation limits emerge, the van der Waals limit for small
separations, and the Casimir limit for large.  In the van der Waals limit
we have
\begin{equation}
 V\left( r \right) \approx - \left( {{{3\hbar } \mathord{\left/ {\vphantom
 {{3\hbar } {2\pi }}} \right.  \kern-\nulldelimiterspace} {2\pi }}}
 \right)\int\limits_0^\infty {d\omega \tilde \alpha _1 \left( {i\omega }
 \right)\tilde \alpha _2 \left( {i\omega } \right)} ;\quad r \ll {c
 \mathord{\left/ {\vphantom {c {\omega _0 }}} \right. 
 \kern-\nulldelimiterspace} {\omega _0 }},
\end{equation}
where $\omega _0 $ is some characteristic frequency above which the
polarizabilities are negligible.  Note that the interaction potential lacks
an $r$-dependence in this range and consequently there is no force.  If we
assume that the characteristic frequencies are the same for the two
composite particles and that the so-called London approximation
\cite{Lon,Lon2}, $\tilde \alpha \left( i\omega
\right) = {{\tilde \alpha \left( 0 \right)} \mathord{\left/ {\vphantom
{{\tilde\alpha \left( 0 \right)} {\left[ {1 + \left( {{\omega
\mathord{\left/ {\vphantom {\omega {\omega _0 }}} \right. 
\kern-\nulldelimiterspace} {\omega _0 }}} \right)^2 } \right]}}} \right. 
\kern-\nulldelimiterspace} {\left[ {1 + \left( {{\omega \mathord{\left/
{\vphantom {\omega {\omega _0 }}} \right.  \kern-\nulldelimiterspace}
{\omega _0 }}} \right)^2 }
\right]}}$, may be used for the polarizabilities the potential in the van
 der Waals limit becomes $V\left( r \right) = - \left( {{3 \mathord{\left/
 {\vphantom {3 8}} \right.  \kern-\nulldelimiterspace} 8}} \right)\tilde
 \alpha _1 \left( 0 \right)\tilde \alpha _2 \left( 0 \right)\hbar \omega _0
 $.  

In the Casimir limit we have
\begin{equation}
 V\left( r \right) = - \left( {{{25\hbar c} \mathord{\left/ {\vphantom
 {{25\hbar c} {32\pi r}}} \right.  \kern-\nulldelimiterspace} {32\pi r}}}
 \right)\tilde \alpha _1 \left( 0 \right)\tilde \alpha _2 \left( 0
 \right);\quad r \gg {c \mathord{\left/ {\vphantom {c {\omega _0 }}}
 \right.  \kern-\nulldelimiterspace} {\omega _0 }},
\end{equation}
and the force is
\begin{equation}
 F\left( r \right) = - \left( {{{25\hbar c} \mathord{\left/ {\vphantom
 {{25\hbar c} {32\pi r^2 }}} \right.  \kern-\nulldelimiterspace} {32\pi r^2
 }}} \right)\tilde \alpha _1 \left( 0 \right)\tilde \alpha _2 \left( 0
 \right).
\end{equation}

Here, the potential has the $1/r$-dependence that we hoped for.  The
polarizability is unit less and very small.  A rough estimate of the transition
point between the two regions is found by equating the Casimir and van der
Waals potentials, using the London approximation for the latter.  Doing so
one finds that $r_c \approx \left( {{{25c} \mathord{\left/ {\vphantom
{{25c} {12\pi \omega _0 }}} \right.  \kern-\nulldelimiterspace} {12\pi
\omega _0 }}} \right)$.  Demanding that this is smaller that the closest
possible distance between the composite particles means that the 
interaction potential may have the $1/r$-dependence at all distances.

\section{Summary and Discussions}\label{Sum}

In summary, we have shown that hypothetical particles with fundamental
interaction potential of harmonic oscillator type lead to retarded
dispersion potentials between composite particles with the same spatial
variation as the gravitational potential and the Coulomb potential.  Both 
these fundamental interactions have serious theoretical problems. In the 
Coulomb case there is the problem with the divergent energy in the fields 
from an electron if it is assumed to be a true point particle. The
present gravity theories have some problems too.  One very serious problem is
that the graviton, the quantum of the quantized gravitational field has
eluded detection.  If one would come to the conclusion that gravity is not
a fundamental interaction after all the present work could come handy.

\begin{acknowledgments}
This research was sponsored by EU within the EC-contract No:012142-NANOCASE
and support from the VR Linn\'{e} Centre LiLi-NFM and from CTS is
gratefully acknowledged.
\end{acknowledgments}


\begin{thebibliography}{10}
\bibitem{Lon}F. London, Zeitschr. f.  Physik. Chem. B{\bf 11}, 222(1930).
\bibitem{Lon2}F. London, Zeitschr.  f. Physik A Hadrons and Nuclei
{\bf 63}, 245(1930).
\bibitem{CasPol} H. B. G. Casimir, and D. Polder, Phys. Rev. {\bf 73}, 360 (1948).
\bibitem{Ser}Bo E. Sernelius, {\it Surface Modes in Physics}, Wiley-VCH (2001).  
\bibitem{Cas} H. B. G. Casimir, J. Chim. Phys. {\bf 47}, 407 (1949).
\bibitem{BosSer} M. Bostr\"{o}m, and Bo E. Sernelius, Phys. Rev. B {\bf 
61}, 2204 (2000).
\bibitem{SerBjo}Bo E. Sernelius, and P. Bj\"{o}rk, Phys. Rev. B {\bf 
57}, 6592 (1998).
\bibitem{Heal}M. A. Heald, and J. B. Marion, {\it Classical Electromagnetic 
Radiation}, Saunders (1995).




\end{thebibliography}
\end{document}